\documentstyle[prc,preprint,eqsecnum,aps]{revtex}

\tighten
\begin{document}
\draft
\preprint{}
\title{Capture of Solar and Higher-Energy Neutrinos by ${}^{127}$I.}
\author{J. Engel}
\address{Dept.  of Physics and Astronomy, CB3255, University of North Carolina,
Chapel Hill, NC 27516 }
\author{S.  Pittel} \address{Bartol Research Institute,
University of Delaware, Newark, DE 19716}
\author{P.  Vogel}
\address{Physics
Department, 161-33, Caltech, Pasadena, CA 91125}
\maketitle
\begin{abstract}

We discuss and improve a recent treatment of the absorption of solar neutrinos
by ${}^{127}$I, in connection with a proposed solar neutrino detector.  With
standard-solar-model fluxes and an in-medium value of -1.0 for the axial-vector
coupling constant $g_A$, we obtain a ${}^8$B-neutrino cross section of
3.3$\times 10^{-42}$, about 50\% larger than in our previous work, and a
${}^7$Be cross section that is less certain but nevertheless also larger than
before.  We then apply the improved techniques to higher incoming energies that
obtain at the LAMPF beam dump, where an experiment is underway to finalize a
calibration of the ${}^{127}$I with electron neutrinos from muon decay.  We
find
that forbidden operators, which play no role in solar-neutrino absorption,
contribute nonnegligibly to the LAMPF cross section, and that the preliminary
LAMPF mean value is significantly larger than our prediction.

\end{abstract} 


\section{Introduction}
\label{sec:intro}

In 1988 Haxton\cite{r:Haxton} proposed ${}^{127}$I as the active ingredient in
a
new solar-neutrino detector.  He estimated that a tank containing 1000 tons of
iodine would detect about 20 times as many neutrinos as the chlorine experiment
at Homestake.  He also noted that a 3/2$^+$ state at 125 keV in ${}^{127}$Xe is
accessible to neutrinos from ${}^7$Be, and that if the ratio of ${}^7$Be and
${}^8$B cross sections were substantially different from that in chlorine, then
the two experiments combined could determine the fluxes of both the ${}^7$Be
and
${}^8$B neutrinos.  Since the ${}^7$Be flux now appears to
be a critical piece of the solar-neutrino puzzle\cite{r:Be7}, a high-statistics
iodine detector has become attractive.

In 1991 we attempted\cite{r:iold} to improve on Haxton's estimate by using the
Quasiparticle Tamm-Dancoff Approximation (QTDA) to calculate the response of
iodine to both ${}^8$B and ${}^7$Be neutrinos.  We concluded that Haxton's
estimate was a little too high, but our results were still sufficiently
encouraging (and carried enough uncertainty) to make a direct calibration
desirable.  In fact, around the same time an attempt was
made\cite{r:Sugarbaker}
to measure the Gamow-Teller (GT) strength distribution, which determines
solar-neutrino cross sections, via the charge-exchange reaction
${}^{127}$I({\it
p,n})${}^{127}$Xe.  Unfortunately, in cases for which no states in the final
nucleus are accessible by beta decay, the reaction determines the GT
distribution only up to an overall normalizing constant.  The solar neutrino
cross section in ${}^{127}$I can therefore not be extracted without measuring
at
least one more quantity --- for example, the absolute GT strength to a
particular state, or the total integrated strength below a certain energy.

To fix the normalization of the GT distribution, and to demonstrate that the
counting of ${}^{127}$Xe works as expected, a group working at LAMPF has
recently exposed an iodine target to a known flux of electron neutrinos from
muon decay\cite{r:Lande}.  The underlying idea is to use the total LAMPF cross
section to fix the unknown constant multiplying the GT distribution.  This
task,
unfortunately, is not completely straightforward.  Because the LAMPF neutrinos
have on average an energy of about 30 MeV, their wavelengths are short enough
to
change the form of the cross section; forbidden operators that depend on
nucleon
coordinates and momenta can modify the ``allowed'' strength from the GT
operator
$\sum_i \sigma_i \tau^+_i$.  For the experiment to yield a reliable
normalization, the forbidden contributions must either be small or accurately
calculated and removed.  The current state of affairs has prompted us to
reexamine neutrino absorption by ${}^{127}$I.  Here, after reviewing our prior
work in Sec\ \ref{s:prior} and developing an improved version in Sec.\
\ref{s:new}, we calculate the LAMPF cross section, including the forbidden
corrections.  These turn out to be large enough to complicate an accurate
extraction of the allowed component from experimental data.

\section{Review of prior work}
\label{s:prior}

In Ref.\ \cite{r:iold} we developed an approximation for odd-mass
nuclei that is closely related to the usual even-even QTDA.  The starting
point is the assumption that the ground state of  ${}^{127}$I is predominantly
of one-quasiparticle character,
\begin{equation}
\pi^{\dag}_{d_{5/2}} \left| BCS \right\rangle ,
\end{equation}
where $\left| BCS \right\rangle $ is the fully-paired BCS quasiparticle vacuum
and
$\pi^{\dag}_{d_{5/2}}$ creates a proton quasiparticle in the $1d_{5/2}$ orbit.
We confirmed our assumption in part by adding three-quasiparticle states of the
form
\begin{equation}
\label{e:istates}
\left[ \left( \pi^{\dag}_i \nu^{\dag}_k \right)^K \nu^{\dag}_l
\right] ^{5/2} \left| BCS \right\rangle  ,
\end{equation}
the admixtures of which turned out to be small.  (In the above equation {\it
i,k,l} are valence orbitals and $\nu^{\dag}_l$ creates a quasineutron in
orbit $l$.)  The form of the iodine ground state implies that the space of
states in xenon accessible via neutrino absorption is largely spanned by the
analogous set
\begin{equation}
\label{e:xestates}
\nu^{\dag}_j \left| BCS \right\rangle  , ~~~~~~~ \left[ \left( \pi^{\dag}_i
\nu^{\dag}_k \right)^K \pi^{\dag}_l \right] ^J \left| BCS \right\rangle  ,
\end{equation}
where $J^{\pi}$ and $j^{\pi}$ assume the values $3/2^+, 5/2^+$, and $7/2^+$,
and $K$ is any intermediate angular momentum.  We therefore used the set
Eq.\ \ref{e:xestates} in calculating the ${}^{127}$Xe spectrum and the GT
strength distribution.

Our valence space for both protons and neutrons consisted of the $2s-1d-0g$
oscillator shell plus the two $0h$ orbitals from the next oscillator shell.  We
took the two-body interaction from Ref.\ \cite{r:Hot} and modified it by
scaling
the pairing matrix elements to reproduce empirical pairing gaps and replacing
the neutron-proton monopole-monopole component with a constant average
interaction.  We changed the monopole component because it alters
single-particle energies through mean field effects, which we accounted for
phenomenologically by taking our single-particle energies from a Wood-Saxon
potential with parameters appropriate for ${}^{127}$I.  The end result of our
calculation was (assuming standard-solar-model fluxes) a ${}^8$B cross section
of 2.2$\times 10^{-42}$ (13 SNU) --- smaller than Haxton's estimate by almost a
factor of three --- and a ${}^7$Be cross section of $2.0 \times 10^{-45} {\rm
cm}^2$ (9.4 SNU).  The ratio of ${}^7$Be to ${}^8$B cross sections was much
larger than in chlorine.

There were of course drawbacks in our approach.  First of all, the QTDA is
number-nonconserving; fluctuations in particle number introduce some error,
which we were unable to estimate.  In addition, we omitted basis states that
might prove important despite the argument above.  For example,
three-quasineutron configurations in ${}^{127}$Xe should combine with the
states
in Eq.\ \ref{e:xestates} to create a low-lying collective $2^+$ phonon, which
could in turn modify the strength distribution at low excitation energies.  In
${}^{127}$I, a quadrupole phonon formed from three-quasiproton configurations
plus the states in Eq.\ \ref{e:istates} could significantly dilute the
one-quasiparticle content of the ground state.  Furthermore, our two-body
interaction was not entirely consistent with our selection of single-particle
energies because we retained the like-particle monopole component while arguing
that the neutron-proton monopole interaction ought to be excluded.  Both can
modify single-particle energies and, moreover, neither is trustworthy no matter
how the energies are chosen (for a convincing demonstration see Ref.\
\cite{r:Zucker}).  Finally, we took no account of spreading widths associated
with our states in xenon.  In revisiting our calculation we have addressed all
these issues.

\section{Improved approach}
\label{s:new}

We have now improved our treatment of neutrino absorption by ${}^{127}$I in
several ways.  To determine the error introduced by number-nonconservation in
the QTDA we have recalculated the GT distribution in the number-conserving
Generalized-Seniority (GS)\cite{r:GS} approximation (which is equivalent to
QTDA
in all other respects).  The one- and three-quasiparticle configurations above
correspond to well-defined states in the GS
approach.  The comparison will be discussed in detail shortly.

We have also made several improvements to the effective force.  For
example we have now removed the monopole component of the interaction both in
the neutron-proton and like-particle channels, and have not replaced them with
anything else.  The average interaction that we included earlier has no effect
(except on binding energies) in the number-conserving GS approach and so can be
omitted.  We argued previously that some remnant of the monopole force might be
needed in a QTDA treatment, where because of the mixing of particle numbers
even
a constant interaction can change wave functions and energies.  We now find,
however, that the effects of number-nonconservation are minimized when the
monopole force is dropped altogether.  In addition, we no longer scale the
like-particle interaction in the pairing channels.  Though the scaling improves
agreement with pairing gaps it badly affects properties of low-lying states
(e.g. energies and magnetic moments).

Beside altering the two-body interaction, we have also changed the way we
select single-particle energies.  In our earlier work, we used the energies
given in Ref.\ \cite{r:Klapdor}; here we take them instead from the spectra of
N=81 nuclei (for neutrons) and Z=51 nuclei (for protons) that are close to
${}^{127}$I and ${}^{127}$Xe.  The new single-particle energies yield a better
ordering of low-lying levels in both mass-127 nuclei --- the old ones reversed
the order, e.g., of the $5/2^+$ and $7/2^+$ levels in ${}^{127}$I.  Another
important change is the inclusion in the QTDA of the three-quasiproton and
three-quasineutron configurations discussed above.  And finally, we have now
taken some account of the background of even more complex states by giving the
final states a spreading width.  Below we discuss all of these changes and
their effects in detail.

In the GS approximation the BCS vacuum is replaced by a number-conserving
involving a condensate of $J=0$ $S$-pairs for
protons and a corresponding condensate for neutrons, viz:
\begin{eqnarray}
\label{e:w=0}
\left| w_{p}  = w_{n}=0 \right\rangle  & \equiv &\left| N_p,N_n \right\rangle
\\
                & = & (S^{\dag}_{p})^{N_{p}} (S^{\dag}_{n})^{N_{n}} \left| 0
\right\rangle  ,
 \nonumber
\end{eqnarray}
where ($\rho = p,n$)
\begin{equation}
\label{e:spair}
S^{\dag}_{\rho} = \sum_j (2j+1)^{\frac{1}{2}}
\alpha_j^{\rho} \left[ \rho^{\dag}_j \rho^{\dag}_j \right]^0
\end{equation}
createsa coherent $J=0$ pair, the state $\left| 0 \right\rangle $ contains no
valence particles,
$N_p$ and $N_n$ are the number of proton and neutron valence $S$ pairs,
$w_{\rho}$ denotes the generalized seniority for particles of type $\rho$,and
the
$\alpha_j^{\rho}$ are (variationally-determined) structure constants related to
the $u_j$'s and $v_j$'s of the BCS formulation.  In this framework, our
original
QTDA treatment of xenon corresponds to diagonalizing the
shell-model hamiltonian in the space of
states
\begin{eqnarray}
\label{e:gsstates}
\left| w_p=0,w_n=1 \right\rangle   & = & \rm{n}^{\dag}_j \left| N_p,N_n
\right\rangle   \\
\left| w_p=2,w_n=1 \right\rangle   & = & \left[ \left( \rm{p}^{\dag}_i
\rm{p}^{\dag}_l \right)^K
\rm{n}^{\dag}_k \right] ^J \left| N_{p} - 1,N_{n}  \right\rangle   , \nonumber
\end{eqnarray}
where $\rm{n}^{\dag}_j$ and $\rm{p}^{\dag}_j$ now create {\it real} particles.
An analogous construction yields the states in iodine corresponding to those in
Eq.\ \ref{e:istates}. (The one spurious state in each nucleus with $K=0$ is
easily removed.)

In Table \ref{t:energies} we display the energies of low-lying states,
alongside
those calculated in the GS and QTDA approximations (without
three-like-quasiparticle states or their GS equivalents).  Neither calculation
does terribly well, but two facts are important.  First, the two agree well
with
one another.  Second, the calculations agree worst with experiment when
three-quasiparticle configurations are important.  This problem will partly be
remedied later when we add the three-like-quasiparticle states.

Fig.~\ref{f:str1} shows the Gamow-Teller strength up to 8 MeV --- the strength
above 7.23 Mev, the neutron-emission threshold is irrelevant for {\it
radiochemical} neutrino detection --- calculated in the same two schemes.
Here,
as in Ref.\ \cite{r:iold} we have reduced the strengths by a phenomenological
quenching parameter (1.0/1.26)$^2$.  The validity of this simple prescription,
which in weak interactions corresponds to setting $g_A$ to -1.0, is not
universally accepted and we will return to it later.  For now we note the
remarkable agreement between the two methods.  The total strength below
threshold is 4.38 in the GS calculation and 4.13 in the QTDA.  The ${}^8$B
cross
sections are 4.39$\times 10^{-42}$ and 3.91$\times 10^{-42}$, respectively,
close to one another but also markedly larger than that reported in Ref.\
\cite{r:iold} (the largest part of the change is due to the pairing matrix
elements, which we no longer renormalize).  The only important disagreement
between the two calculations is in the strength to the lowest $3/2^+$ state:
.079 in the GS scheme and .042 in the QTDA.  Though it is unreasonable to
expect
calculations designed to reproduce the entire spectrum to match up state by
state, in this instance the disagreement is disappointing because this one
strength completely determines the cross section for ${}^7$Be neutrinos.  The
GS
strength is more than twice as large as in our previous paper (.035) and if
correct would (again with standard solar fluxes) result in more events from
${}^7$Be than from ${}^8$B.  By comparison the QTDA result for the first
$3/2^+$
state is only marginally bigger than before.  The GS calculation is clearly
preferable since it does not violate particle number, but because of its
complexity we have not extended it to include more configurations (which can
also affect the ${}^7$Be strength), as we have in the QTDA.  We therefore
cannot
make a terribly strong statement about the ${}^7$Be cross section; all signs,
however, point to its being much larger with respect to the ${}^8$B cross
section than in ${}^{37}$Cl.  Other properties are very similar in the two
approaches provided all monopole forces are set to zero.  For the rest of this
paper, then, we will no longer use the GS scheme, since the QTDA is
computationally much simpler to extend.

We next examine the effects of the three-quasineutron and three-quasiproton
configurations in the QTDA.  Table \ref{t:energies1} shows the same energy
levels as in Table \ref{t:energies}, but with the extra configurations
included.
The result is a dramatic lowering in energy of some of the mainly
three-quasiparticle states, reflecting the formation via the neutron-proton
interaction of a collective quadrupole phonon.  The new configurations also
increase the density of low-lying states substantially.  Though the
quantitative
agreement in the lowest states is still not impressive, these new results are a
considerable improvement.  Furthermore, as noted above, the method is designed
to encompass the entire final-state spectrum (in our model space several
thousand states); the lowest lying states, with the exception of the first
$3/2^+$, are no more important than the others.

The formation of phonon-like states shifts the E2 strength downwards, although
no single multiplet is as collective as implied by experiment.  This may be an
indication that further correlations associated with higher quasiparticle
number
(or with deformation) ought to be included at some level.  The same conclusion
might be drawn from the ${}^{127}$I ground-state magnetic moment, which changes
(with free g-factors) from 4.39 to 3.76.  (The experimental value is 2.81 and
the Schmidt value is 4.79.)  The incomplete but still significant improvement
implies that spin correlations, important for parts of the GT distribution, are
represented better than before, though again not perfectly.  It also suggests
that at least some extra quenching of the spin operator is in order.

In Fig.~\ref{f:str2} we compare the QTDA GT strength below the neutron-emission
threshold with and without the three-like-quasiparticle configurations.  The
difference, while nonnegligible, is not dramatic.  The low-lying phonon-like
states apparently do little to the charge-changing strength, which is in a
different channel.  The main effect, a small overall loss of strength is due
largely to the reduction of the single-quasiparticle component of the
${}^{127}$I ground state from 86\% to 76\%.  The ground state
three-quasiparticle components go mainly into {\it five}-quasiparticle states
in
${}^{127}$Xe, which are not included in our basis.  To the extent that these
states lie below the neutron-emission threshold (they should begin at roughly 4
times the pairing gap --- about 4-5 MeV), we will underestimate the ${}^8$B
cross section.  Phase space considerations reduce the importance of states this
high in energy, however; in this application, therefore, the error should be
relatively small.

The final new element in this paper is the inclusion of spreading widths.
These
depend on the density of complicated background states and on the average
interaction strength coupling them to our model space, neither of which we can
reliably estimate.  We settle instead for the prescription in Ref.\
\cite{r:widths}, developed primarily to describe spreading of RPA-like states
high in energy, viz:
\begin{eqnarray}
\label{e:width}
\Gamma(\omega) & \approx  & \frac{1}{\omega} \int_0^{\omega} d\epsilon \,\left[
\gamma(\epsilon) + \gamma(\epsilon - \omega) \right] \\
\gamma(\epsilon) & = & 10.75 \left( \frac{\epsilon^2}{\epsilon^2 + 18^2}
\right)
\left( \frac{110^2}{\epsilon^2 + 110^2} \right),  \nonumber
\end{eqnarray}
where $\omega$ is the excitation energy, $\gamma(\epsilon)$ is a
parameterization of the single-quasiparticle width (in MeV), and we have
included a lower limit of 300 keV to simulate experimental resolution.
Fig.~\ref{f:spread} shows the full GT distribution with the widths included
(and
divided by .76 {\it in the figure only} to account for the strength in
five-quasiparticle states, so that the height of the giant resonance is roughly
correct).  The spreading turns out to have a fairly small effect on both the
total strength below threshold and the ${}^8$B cross section.  Our final values
for these quantities, with all the new physics included and the spin operator
quenched as discussed above, are 3.00 and 3.27$\times 10^{-42}$.  The ${}^8$B
cross section is 50\% larger than our old value.  Using the free-nucleon
axial-vector coupling constant would increase it another 60\% and we can take
that figure as a nominal measure of uncertainty in our result.

\section{Higher-energy neutrinos and forbidden corrections}
\label{s:LAMPF}

Our strength distribution does not differ dramatically from the one extracted
from experiment\cite{r:Sugarbaker}.  Both suffer from uncertainty in the
overall
normalization, which, as discussed in the introduction, has prompted an attempt
at LAMPF\cite{r:Lande} to measure the total cross section for absorbing
electron
neutrinos from muon-decay.  But the energies of these neutrinos extend to half
the muon mass, much higher than the solar neutrino end point ($\approx 14$
MeV),
and the momenta transferred are typically on the order of the inverse radius of
the target nuclei.  The allowed approximation, which ignores variations
in the lepton wave functions over the interior of the nucleus, no longer
applies
and the cross section depends on nuclear wave functions in a more complicated
way.  Ref.\ \cite{r:Walecka} contains a framework for treating neutrino-induced
reactions beyond the allowed approximation; it involves calculating matrix
elements of the operators:
\begin{eqnarray}
\label{e:ops}
j_J(qr) Y_J(\Omega_r) \nonumber \\
j_L(qr) [Y_L(\Omega_r) \, \vec{\sigma} ]^J \nonumber  \\
j_L(qr) [Y_L(\Omega_r) \, \frac{1}{M} \vec{\nabla} ]^J \nonumber \\
j_J(qr) Y_J(\Omega_r) \, \vec{\sigma} \cdot \frac{1}{M} \vec{\nabla}
\end{eqnarray}
where $q$ is the momentum transfer, $j_L(qr)$ is the $L$th spherical Bessel
function, $Y_L(\Omega_r)$ is a spherical harmonic, $J$ is the multipole order
of
the
transition, and $M$ is the nucleon mass.  The maximum value of $qr$ is
about 2.8 and we need not consider multipoles above $J \approx 3$.

With our QTDA wave functions the evaluation of the beam-dump cross section is
nominally straightforward, but a few subtleties complicate the analysis.  The
most important difficulty arises in connection with the Coulomb attraction of
the outgoing electron towards the $Z=54$ xenon nucleus.  In both the low-energy
($qr \ll 1$)\cite{r:Fermifunction} and high-energy ($qr \gg
1$)\cite{r:Rosenfelder} limits, clear procedures exist for evaluating
Coulomb effects; in one case the outgoing electron is known to be in an
$s$-state, enhancing the cross section by the usual Fermi factor, and in the
other the outgoing wave function can be treated as a plane wave perturbed
slightly by the charge distribution in the nuclear interior.  Since $qr \approx
1$ here, neither limit applies and so we follow a different line of
reasoning.

The multipolarity of the nuclear transition must be equal to that obtained from
coupling the incoming and outgoing lepton angular momenta and parities.  As we
shall show shortly, the most important transitions have multipole quantum
numbers $1^+$ and $2^-$.  For $1^+$ transitions, the orbital angular momenta of
the two leptons must be coupled to either 0, or 2.  The latter possibility
requires both the neutrino and the electron to have $l = 1$ (or for one of them
to have $l = 2$), which for the energies of interest here is less likely than
the other possibility --- that they both have $l=0$, i.e.  that they are both
in
$s$-states.  We therefore treat Coulomb effects in the $1^+$ channel as if the
transitions were purely allowed --- that is, we multiply the $1^+$ cross
section
by the usual Fermi function $F(Z,E)$.

The $2^-$ transitions require at least one lepton to be in a $l=1$
(or higher) state, and the usual prescription makes less sense.  These are
``unique" transitions, however, and can be treated simply in processes such as
beta decay\cite{r:Schopper,r:Behrens}, for which $qr$ is always small.  The
procedure there is to modify the usual cross section (with Fermi function) by a
factor
\begin{equation}
\label{2^-} \frac{p_{\nu}^2 + \lambda_2 p_e^2}{p_{\nu}^2 + p_e^2} ,
\end{equation}
where $p_{\nu}$ and $p_e$ are the neutrino and electron momenta, and
$\lambda_2$
is a function of $p_e$ tabulated in Ref.\ \cite{r:Fermifunction}.  Though $qr$
is not always small enough in our problem for the procedure to be strictly
valid, we employ it to get a rough handle on the Coulomb corrections; they turn
out to add very slightly to the corrections produced by the Fermi function.

Table \ref{t:multipole} shows the contribution of each multipole to the
beam-dump cross section, evaluated with the two commonly used values for $g_A$
(-1.0 and -1.26).  Several effects emerge.  The presence of the Bessel
functions
in the second, third, and fourth operators in Eq.\ \ref{e:ops}, and of $q/M$ in
the third and fourth operators, together reduce the $1^+$ contribution to about
2/3 of its allowed value.  (The Bessel-function effects can be computed
reliaby;
in fact, the same calculation in the simple Helm model\cite{r:Helm} gives a
very
similar result.)  The contributions of the $2^+$ and $3^+$ operators, in which
$j_2$ is the first Bessel function to appear, are nominally suppressed by the
upper limit in $qr$ but are in fact far from negligible.  The $0^+$, on the
other hand, contributes little even though it is unsupressed because its
strength is concentrated in the isobar analog state, which lies several MeV
above the neutron-emission threshold.  The $0^-$ and $1^-$ strengths are even
more concentrated --- within the giant-dipole resonance and its spin analog,
well above 7.23 MeV.  In fact the $2^-$ contribution, which we find to be of
the
same order as that of the $1^+$, may well be smaller than in our table, because
the associated strength is probably also concentrated at higher energies.  Our
model space is clearly not large enough for the full collectivity of the
spin-dipole mode to assert itself.

To get some idea of the size of our overestimate we performed QRPA calculations
for ${}^{128}$Xe with the same single-particle levels and effective interaction
we used in the odd-mass nuclei.  We then enlarged the model space to include
the
$1f$ and $2p$ levels a few MeV above the Fermi surface.  We found that while
the
total $2^-$ strength increased, the amount below 7 MeV fell by 20 -- 30\%.
Including other levels below the Fermi surface (e.g.  the $0f-1p$ shell) would
concentrate it further still; just how much would remain below threshold is not
clear.

Altogether, the cross section for muon-decay neutrinos is only about half of
the
preliminary experimental value of 6$\times 10^{-40}$ cm$^2$ \cite{r:Lande},
even
without any quenching of the axial current.  There is no clear way to make our
results compatible with this larger cross section.  As is apparent from the
table, the bulk of the cross section comes from the $1^+$ and $2^-$ multipoles,
and we have argued that the $2^-$ contribution is probably an overestimate.
Because the average neutrino energy is so much higher than 7 MeV, phase space
plays only a small role and the Gamow-Teller $1^+$ contribution depends
essentially only on the total strength below the neutron-emission threshold ---
it is insensitive to the precise distribution.  To account for the difference
between our calculated total cross section and a value of 6$\times 10^{-40}$
cm$^2$, the $1^+$ contribution would have to be about three times larger than
in
Table \ref{t:multipole}.  An increase of about 25\% might be plausible since,
as
discussed in connection with Fig.\ \ref{f:spread}, we have not included states
with five or more quasiparticles.  Our claim, however, is that such states
appear in large numbers only at energies above 4 -- 5 MeV and that relatively
few will contribute to the strength below threshold.  In our calculation the
summed GT strength below threshold is about 10\% of the total strength, which
is
dominated ($\approx$ 70\%) by the states that form the giant Gamow-Teller
resonance at 13-14 MeV.  Such a distribution is consistent with general
experience in other nuclei; in fact it seems impossible to increase the
low-lying GT strength substantially as long as the giant GT resonance is at the
energy suggested by the $(p,n)$ reaction, which in turn is in reasonable
agreement with our calculation.  Finally, though it is not impossible that we
are seriously underestimating the contributions of the other multipoles, that
would not be good news either.  A forbidden contribution even larger than our
estimate would make an extraction of the GT distribution extremely difficult.
Unfortunately, therefore, we cannot reconcile our results with the value
6$\times 10^{-40}$cm$^2$ without invoking an unforeseen mechanism that would
probably spoil the calibration.

\section{Conclusions}
\label{s:conclusion}

In summary, we have recalculated cross sections for the capture by ${}^{127}$I
of neutrinos from solar ${}^8$B and ${}^7$Be decay --- incorporating several
new
physical effects --- and have found somewhat larger values than in our previous
work.  The physics we still have not included, e.g.  further
correlations/deformation, a precisely determined and justified value for $g_A$,
a more carefully crafted interaction, etc., make our results somewhat
uncertain,
but with enough effort more comprehensive calculations are possible.  Our
current results, we feel, still support the development of a calibrated
iodine-based solar neutrino detector if only because it would be an improved
version of the existing Homestake experiment, which as of now is the most
difficult to reconcile with standard-model physics.  We must also conclude,
unfortunately, that a complete calibration may be more difficult than
originally
hoped.  We do not see how the preliminary LAMPF cross section can be due
entirely to Gamow-Teller-like strength, and do not as yet know how to evaluate
and remove the forbidden contributions with the required accuracy.
Nonetheless,
we are hopeful that with the steady advance of experimental techniques,
nuclear-structure expertise, and computing power, the remaining problems can be
overcome.

\acknowledgments

We wish to acknowledge useful conversations with W.C. Haxton and K. Lande. This
work was supported in part by the U.S. Department of Energy under grants
DE-FG05-94ER40827 and DE-FG03-88ER40397, and by the National Science Foundation
under grants PHY-9108011 and PHY-9303041.

\begin{figure}
\caption{The Gamow-Teller strength B(GT) (quenched and in half-MeV bins) from
${}^{127}$I to states below 8 MeV in ${}^{127}$Xe in the GS scheme (solid line)
and in the QTDA (dashed line).}
\label{f:str1}
\end{figure}

\begin{figure}
\caption{The Gamow-Teller strength B(GT) (quenched and in half-MeV bins) from
${}^{127}$I to states below 8 MeV in ${}^{127}$Xe in the QTDA, with
three-like-quasiparticle configurations included (solid line), and without them
(dashed line --- same as in Fig.~1).}
\label{f:str2}
\end{figure}

\begin{figure}
\caption{The Gamow-Teller strength distribution B(GT) per 100 keV (quenched)
from ${}^{127}$I in the QTDA, with
three-like-quasiparticle configurations and spreading widths included.  The
curve, which here goes up to 20 MeV in
${}^{127}$Xe, has been scaled by 1/.76 to account for missing
five-quasiparticle
states lying largely in the giant resonance.}
\label{f:spread}
\end{figure}

\begin{table}
\caption{Energies of low-lying states in ${}^{127}$I and ${}^{127}$Xe in keV.
The second column contains measured values, the third the QTDA predictions, and
the last the GS predictions.  Neither calculation
incorporates three-like-quasiparticle states (or their analogs in the GS
scheme).
\label{t:energies}}
\begin{tabular}{ccrrr}
& $J^{\pi}$ & exp. & ~~QTDA$ $ & GS  \\
\tableline
${}^{127}I$  & & & & \\
 & 5/2${}^+$ &      0  &            0   &           0 \\
 & 7/2${}^+$ &     58  &          -14   &         62 \\
 & 3/2${}^+$ &    203  &          1261  &        1407  \\
 & 1/2${}^+$ &    375  &          1171  &        1299  \\
 & 5/2${}^+$ &    418  &          1695  &        1929 \\
${}^{127}Xe$  & & & & \\
 & 1/2${}^+$ &    0    &       0    &         0  \\
 & 3/2${}^+$ &   125   &     178    &       300  \\
 & 9/2${}^-$ &   297   &    1449    &      1590  \\
 & 11/2${}^-$&   309   &     -16    &        80  \\
 & 3/2${}^+$ &   322   &    1351    &      1430  \\
 & 7/2${}^+$ &   346   &     853    &       860  \\
 & 5/2${}^+$ &   376   &    1009    &       970  \\
 & 1/2${}^+$ &   412   &    1530    &      1430  \\
 & 5/2${}^+$ &   510   &    1428    &      1560  \\
 & 3/2${}^+$ &   587   &    1616    &      1720  \\
 & 9/2${}^+$ &   646   &    1588    &      1630  \\
\end{tabular}
\end{table}

\begin{table}
\caption{Energies of low-lying states in ${}^{127}$I and
${}^{127}$Xe in keV.  The second
column reprises the measured energies and the third column contains the
predictions of the QTDA with three-like-quasiparticle configurations included.
They should be compared with the results of the simpler QTDA calculation in
Table 1.
\label{t:energies1}}
\begin{tabular}{cccc}
& $J^{\pi}$ & exp. & QTDA (improved) \\
\tableline
 ${}^{127}$I & &  & \\
& 5/2${}^+$ &      ~~0  &        ~~~0  \\
& 7/2${}^+$ &     ~58  &       ~~67   \\
& 3/2${}^+$ &    203  &      ~636  \\
& 1/2${}^+$ &    375  &      ~941  \\
& 5/2${}^+$ &    418  &      ~555   \\
 ${}^{127}$Xe  & & & \\
& 3/2${}^+$ &   125   &     ~216   \\
& 9/2${}^-$ &   297   &     ~360   \\
& 11/2${}^-$&   309   &     ~153   \\
& 3/2${}^+$ &   322   &     ~547   \\
& 7/2${}^+$ &   346   &     ~630   \\
& 5/2${}^+$ &   376   &     ~367   \\
& 1/2${}^+$ &   412   &     ~898   \\
& 5/2${}^+$ &   510   &     ~664   \\
& 3/2${}^+$ &   587   &     ~682   \\
& 9/2${}^+$ &   646   &    1528   \\
\end{tabular}
\end{table}

\begin{table}
\caption{Contributions of individual multipoles to the total cross section for
neutrinos from muon decay, in units of $10^{-40} {\rm cm}^2$.  The two columns
correspond to correspond to commonly used values for $g_A$ (see text).
\label{t:multipole}}
\begin{tabular}{ccc}
$J^{\pi}$ & $g_A = -1.0$ & $g_A = -1.26$ \\
\tableline
 0${}^+$ &   0.096     &  0.096 \\
 0${}^-$ &   0.00001   &  0.00002  \\
 1${}^+$ &   1.017     &  1.528  \\
 1${}^-$ &   0.006     &  0.008 \\
 2${}^+$ &   0.155     &  0.213  \\
 2${}^-$ &   0.693     &  1.055 \\
 3${}^+$ &   0.149     &  0.171  \\
 3${}^-$ &   0.017     &  0.025 \\
\tableline
total    &   2.098     &  3.096 \\
\end{tabular}
\end{table}

\end{document}